\documentclass{article}
\input{tcilatex}
\begin{document}

\begin{center}
\bigskip

\bigskip ${\LARGE THE\ CGLMP\ BELL\ INEQUALITIES}$\bigskip

B. J. Dalton$^{1,2}$

\bigskip

$^{1}$Centre for Quantum Science and Technology Theory, Swinburne University
of Technology, Melbourne, Victoria 3122, Australia

$^{2}$School of Physics and Astronomy, University of Glasgow, Glasgow G12
8QQ, United Kingdom

\bigskip
\end{center}

*Correspondence: bdalton@swin.edu.au\bigskip

\textbf{Abstract: }Quantum non-locality tests have been of interest since
the 1960's paper by Bell on the original EPR paradox.The present paper
discusses whether the CGLMP (Bell) inequalities obtained by Collins et al
are possible tests for showing that quantum theory is not underpinned by
local hidden variable theory (LHVT). It is found by applying Fine's theorem
that the CGLMP approach involves a LHVT for the joint probabilities
associated with the measurement of one observable from each of the two
sub-systems, even though the underlying probabilities for joint measurements
of all four observables involve a hidden variable theory which is not
required to be local. The latter HVT probabilities involve outcomes of
simultaneous measurements of pairs of observables corresponding to
non-commuting quantum operators, which is allowed in classical theory.
Although the CGLMP inequalities involve probabilities for measurements of
one observable per sub-system and are compatible with the Heisenberg
uncertainty principle, there is no unambiguous quantum measurement process
linked to the probabilities in the CGLMP\ inequalities. Quantum measurements
corresponding to the different classical measurements that give the same
CGLMP\ probability are found to yield different CGLMP\ probabilities.
However, violation of a CGLMP inequality based on any one of the possible
quantum measurement sequences is sufficient to show that the Collins et al
LHVT does not predict the same results as quantum theory. This is found to
occur for a state considered in their paper - though for observables whose
physical interpretation is unclear. In spite of the problems of comparing
the HVT inequalities with quantum expressions, it is concluded that in spite
of the contextuality loophole the CGLMP inequalities are indeed suitable for
ruling out local hidden variable theories\ and also non-local ones as well.
The state involved could apply to a macroscopic system, so the CGLMP\ Bell
inequalities are important for finding cases of macroscopic violations of
Bell locality. Possible experiments in double-well Bose condensates
involving atoms with two hyperfine components are discussed. \bigskip

\textbf{Keywords}: Hidden variable theory, Quantum non-locality, Bell
inequalities, Copenhagen quantum interpretation, Localism, Contextuality

\begin{center}
\pagebreak
\end{center}

\section{Introduction}

\label{Section - Introduction}

The concept of hidden variable theory was introduced in papers by Einstein,
Schrodinger, Bell and Werner (\cite{Einstein35a}, \cite{Schrodinger35a}, 
\cite{Schrodinger35b}, \cite{Bell65a}, \cite{Werner89a}). Einstein suggested
that quantum theory, though correct was \emph{incomplete} - in that the 
\emph{probabilistic} measurement outcomes predicted in quantum theory could
be just the \emph{statistical }outcome of an underlying \emph{deterministic}
theory, where the possible measured outcomes for all observables \emph{always%
} have specific values, and measurement merely reveals what these values
are. Hence observable quantities (such as position and momentum) could be
regarded as elements of \emph{reality} irrespective of whether an \emph{%
actual} measurement has taken place. In contrast, in the Copenhagen
interpretation of quantum theory (see \cite{Copenhagen} for a discussion),
the values for observables do not have a presence in reality\emph{\ until}
measurement takes place. The EPR paradox challenged this viewpoint and
involved an entangled state for two \emph{well-separated} and no longer
interacting distinguishable particles, which had well-defined values for the
position \emph{difference} and the momentum \emph{sum}. For this state,
measuring the position (or the momentum) for the first particle would affect
the outcome for measuring the position (or the momentum) of the second
particle (a feature we now refer to as \emph{steering}). As this was to
happen \emph{instantaneously}, Einstein regarded this as being in conflict
with \emph{causality}. Furthermore, by measuring the position for the first
particle, the \emph{position} for the second particle would then be known 
\emph{without} doing a measurement on the second particle - because the
position difference was well-defined. The paradox is that by then measuring
the\emph{\ position} of the first particle and\emph{\ }the\emph{\ momentum}
for the second particle, a joint \emph{precise} measurement of \emph{both}
the position and momentum for the second particle would have occurred - in
conflict with the Heisenberg uncertainty principle. The Schrodinger cat
experiment \cite{Schrodinger35b} is another paradox, but now involving a 
\emph{macroscopic} sub-system (the cat) in an entangled state with a \emph{%
microscopic} sub-system (the two state radioactive atom). From the Einstein
concept of reality the cat must be \emph{either} alive \emph{or }dead even 
\emph{before} the box is opened to see what is the case. However, from the
Copenhagen viewpoint the cat is \emph{neither} dead \emph{nor} alive \emph{%
until} the box is opened - which conflicts with common-sense. The cat being
neither dead nor alive is a paradox in the Einstein concept of reality
though not from the Copenhagen viewpoint. Bohm \cite{Bohm51a} described a
similar paradox to EPR, but now involving a system consisting of two spin $%
1/2$ particles in a singlet state, and where the observables were spin
components with \emph{quantized} measured outcomes rather than the \emph{%
continuous} outcomes that applied to EPR.\medskip

Einstein believed that an underlying \emph{realist} theory could be found,
based on what are now referred to as\emph{\ hidden variables} - which would
specify the real or underlying state of the system. However, it was not
until 1965 before a quantitative general form for \emph{local} \emph{hidden
variable theory} (LHVT)\ was proposed by Bell \cite{Bell65a}. This was
relevant for the EPR paradox and could be tested in experiments. In its
simplest form, the key idea is that hidden variables are specified
probabilisticly when the state for the composite system is prepared, and
these would determine the \emph{actual} values for \emph{all} the sub-system
observables even after the sub-systems have separated - and even if the
observables were\emph{\ incompatible} with \emph{simultaneous} precise
measurements (such as two different spin components). In the EPR\ experiment
they would specify \emph{both} the position and momentum for each
distinguishable particle. More elaborate versions of local hidden variable
theory only require the hidden variables to determine the \emph{probabilities%
} of measurement outcomes for each sub-system observable, with the overall
expressions for the joint sub-system measurement outcomes being obtained in
accordance with classical probability theory (see \cite{Wiseman07a}, \cite%
{Reid09a}, \cite{Brunner14a}, \cite{Jevtic15a}, \cite{Dalton16a} for a
description). States where the joint probability can be described via local
hidden variable theory are referred to as \emph{Bell local}. Quantum states
for composite systems that could be described by local hidden variable
theory were such that certain inequalities would apply involving the mean
values of products for the results of measuring pairs of observables for
both sub-systems - the \emph{Bell inequalities} \cite{Bell65a}, \cite%
{Bell71a}. States for which a local hidden variable theory does not apply
(and hence violate Bell inequalities) are the \emph{Bell non-local} states.
Based on the entangled singlet state of two spin $1/2$ particles Clauser et
al \cite{Clauser69a} proposed an experiment that could demonstrate a
violation of a Bell inequality\ (CHSH inequality). This would show that
local hidden variable theory could not account for experiments that can be
explained by quantum theory. Subsequent experimental work violating Bell
inequalities confirmed that indeed there are some quantum states for which a
local hidden variable theory does \emph{not} apply and where quantum theory
was needed to explain the results (see Brunner et al \cite{Brunner14a} for a
recent review). The existence of \emph{some} quantum states (such as the two
qubit \emph{Bell states} \cite{Barnett09a}) for which the Bell inequalities
are \emph{not} obeyed and which was confirmed experimentally is itself
sufficient to show that Einstein's hope that an underlying reality
represented by a local hidden variable theory could \emph{always} underpin
quantum theory \emph{cannot} be realised.\medskip

As Brunner et al \cite{Brunner14a} point out, a wide variety of different
Bell inequalities have been derived. A class of Bell inequalities introduced
by Collins et al \cite{Collins02a} (CGLMP) are of particular interest. The
CGLMP inequalities are of particular significance in that they can be
violated for a quantum state for a \emph{macroscopic} system, as Collins et
al show in their paper. The Collins et al \cite{Collins02a} formalism is
based on a HVT in which the fundamental probability $C(j,k,l,m)$ (for which
no quantum expression exists) is for the outcomes of measuring \emph{four}
observables ($A_{1},A_{2}$ for one sub-system, $B_{1},B_{2}$ from the other,
with the outcomes listed as $j,k$ and $l,m$ accordingly), but where the
observables for the same sub-system would be incompatible according to
quantum theory. The authors state that their approach is a\emph{\ local}
hidden variable theory, but it is not obvious why this is the case. This is
because the derivation of the CGLMP\ inequalities does not require the HVT
for the $C(j,k,l,m)$ to be local, and to involve \emph{separate}
probabilities $D(j,k)$ and $E(l,m)$ for the observables of each sub-system -
a key requirement for locality. However, as we point out in the present
paper this is not required for the theory to be local. A theorem due to Fine 
\cite{Fine82a} shows that the \emph{marginal probabilities} associated with
measurements of \emph{one} observable per sub-system $P(\alpha _{a},\beta
_{b}|A_{a},B_{b})$ (where $a,b=1,2$ and where $\alpha _{1}\equiv j,\alpha
_{2}\equiv k,\beta _{1}\equiv l,\beta _{2}\equiv m$) can \emph{always} be
described by a local hidden variable theory, and this is all that is
required to justify their claim. Thus, it is the marginal probabilities $%
P(\alpha _{a},\beta _{b}|A_{a},B_{b})$ that can be expressed in LHVT\ form,
not the $C(j,k,l,m)$. As shown in this paper, the proof of the CGLMP
inequalities does not depend on the HVT probabilities $C(j,k,l,m)$
themselves satisfying the locality requirement. \medskip

Although the validity of introducing in a classical HVT the fundamental
probability $C(j,k,l,m)$ for simultaneous measurements of incompatible
observables is not being challenged\textbf{\ }(except when we consider the
contextuality loophole \cite{Khrennikov08a}, \cite{Nieuwienhuizen11a}, where
the assumed existence of the $C(j,k,l,m)$ is challenged), it does have
consequences. It results in there being no \emph{unambiguous} quantum
measurement process for treating the probabilities involved in the CGLMP\
inequalities. However, it is contended here that in spite of there being no
unambiguous quantum measurement process, comparisons between the HVT and
equivalent quantum predictions are still possible - and these are \emph{%
sufficient} to show that the Collins et al \cite{Collins02a} HVT does not
predict the same results as quantum theory in all cases. \medskip

. The basis for this contention may be summarized as follows: Firstly, in
order to compare the predictions of the Collins et al \cite{Collins02a}
version of HVT with those from quantum theory, the two sets of predictions
must be applied to the outcomes for the \emph{same} measurement processes.
Secondly, so as to avoid an immediate conflict with the Heisenberg
uncertainty principle, the actual measurement processes involved in
determining the quantities in the CGLMP inequalities must \emph{not involve}
simultaneous measurement of observables that are incompatible according to
quantum theory - otherwise there would be no quantum theory expressions
available to determine the relevant probabilities. The quantum measurement
processes considered are projective measurements, with outcome probabilities
as in Eq. (\ref{Eq.QThyOverrProbBothMeasts}) below. The measurement process
considered by Collins et al \cite{Collins02a} avoids conflict with the
Heisenberg uncertainty principle by just involving steps where only \emph{one%
} observable from each sub-system is being measured at a time. In compliance
with this requirement, the CGLMP inequalities involve probabilities for the
outcomes of such pairs of observables, with the outcomes for the other pair
of observable being left unrecorded. However, as a consequence there are 
\emph{several} differing measurement processes that are equivalent in
classical HVT which yield the same probability for the overall measurement
process. These differ in the \emph{order} in which measurements on the two
pairs of sub-system observables is made, and on whether or not measurements
are \emph{actually} made on the pair of observables whose outcomes are left
unrecorded. However, although these differing measurement processes yield
the same final outcome probability in \emph{classical} physics, the same is
not the case in \emph{quantum} physics. Fundamentally this is because
quantum measurements \emph{change} the state whereas classical measurements
do not. Following the Copenhagen interpretation of quantum \ theory, the
density operator changes during the process via quantum projector operators
that correspond to the quantum measurement that has taken place. The new
density operator must of course still satisfy the condition $Tr\widehat{\rho 
}=1$. The density operator also changes when outcomes are left unrecorded.
On the other hand, it does not change if no measurement is made. As we will
see, the quantum theory expressions differ if the pairs of observables are
measured in a different order, and not measuring a pair of observables
yields a different outcome probability for the other pair of observables,
than if the first pair are measured and their outcomes disregarded. Hence
there are \emph{several} different quantum theory expressions that
correspond to the probabilities occurring the CGLMP\ inequalities, a feature
that has not previously been recognized. \medskip

In spite of there being no unique quantum measurement process, then because
each of these quantum measurement processes is equivalent to a classical
measurement process from which the probabilities in the CGLMP inequalities
can be obtained, a \emph{violation} of a CGLMP\ inequality based on \emph{any%
} one of the quantum measurement processes is \emph{sufficient} to show that
the Collins et al \cite{Collins02a} HVT does not predict the same results as
quantum theory. The most convenient quantum measurement process is actually
the one where pairs of observables whose results are to be left unrecorded,
are \emph{never measured} at all. Based on this expression, Collins et al 
\cite{Collins02a} find a quantum state (the maximally entangled state of two 
$d$ dimensional systems) that displays a violation of the CGLMP\ inequality $%
I\leq 3$.\medskip\ 

One issue (apart from the contextuality loophole)\textbf{\ }still remains -
the observables involved for the CGLMP violation presented in \cite%
{Collins02a} have no obvious physical interpretation, as they are associated
with Hermitian operators that are off-diagonal in the basis states for each
sub-system. In principle however, such operators could be measured.
\medskip\ 

So far no experiments confirming a CGLMP Bell violation have been reported.
As Bose-Einstein condensates in cold atomic gases are now available based on
double-well potentials supporting localized modes, cases are available where
there are two localized modes per well associated with different hyperfine
states. With the modes for each well defining two sub-systems, the two
different observables for each sub-system could be defined in terms of the
Schwinger spin states for each sub-system. With equal numbers $N/2$ of
bosons in each sub-system it may be possible to prepare a maximally
entangled state with $d=N+1$. It would be of particular interest to see if
there is a violation of the CGLMP\ inequalities for such a system, since if
the number of bosonic atoms is large, an experimental situation for
confirming general non-locality in a macroscopic system may be available.
\medskip

In Section \ref{Section - CGLMP Formalism} the basic features of the CGLMP
formalism will be reviewed - including the relationship between the
fundamental probability introduced and probabilities that appear in the
CGLMP inequalities. The classical measurement processes associated with such
inequalities is identified. In Section \ref{Section - Local HVT and CGLMP
Inequalities} the relationship between the CGLMP\ formalism and hidden
variable theory (both non-local and local) is described, and one of key
CGLMP\ inequalities is derived. In Section \ref{Section - CGLMP and
Measurements} the issue of replicating the classical measurement processes
associated with the CGLMP\ inequalities with measurement processes for which
there is a quantum theory formalism is treated - including a comparison with
the standard local HVT situation where the fundamental probability
introduced only involves one observable for each sub-system. The quantum
theory expressions used to determine the quantities in the CGLMP\
inequalities are identified, and a quantum state for which an inequality
violation occurs is referred to. The contextuality loophole is discussed in
Section \ref{Section - Contextuality Loophole}.\textbf{\ }Section \ref%
{Section - Summary} summarizes the results. Details are set out in the
Appendix.\medskip

In this paper the same symbols will be used for the measurement outcomes,
but classical HVT observables will generally be distinguished from quantum
observables by the absence of the operator symbol. Quantum theory
probability expressions will have a subscript $Q$. \bigskip

\section{The CGLMP Formalism}

\label{Section - CGLMP Formalism}

In this section the fundamental HVT\ probabilities $C(j,k,l,m)$ for joint
measurements of \emph{two} sub-system observables for \emph{each} of the two
sub-systems introduced by Collins et al \cite{Collins02a} are seen as
describing measurement outcomes possible in classical physics, though not in
quantum physics. The issue of whether the HVT is local or non-local will be
examined. It can be seen that all of the probabilities for joint
measurements of \emph{one} sub-system observable for each of the two
sub-systems $A$, $B$ of a bipartite system introduced by Collins et al \cite%
{Collins02a} are also recognizable as standard probabilities in classical
physics. All can be validly expressed in terms of the fundamental joint
measurement probabilities $C(j,k,l,m)$. \medskip\ 

\subsection{Probabilities introduced by Collins et al}

In the standard notation we would express the probability $C(j,k,l,m)$ in
Collins et al \cite{Collins02a} that the measurement of observables $%
A_{1},A_{2},B_{1},B_{2}$ results in outcomes listed as $j,k,l,m$ as 
\begin{equation}
C(j,k,l,m)\equiv P(j,k,l,m|A_{1},A_{2},B_{1},B_{2})
\label{Eq.CollinsBasicProb}
\end{equation}%
Here the observables for sub-system $A$ are $A_{1},A_{2}$ and those for
sub-system $B$ are $B_{1},B_{2}$, and all four observables have the \emph{%
same number} $d$ of different outcomes - listed as $j,k,l,m=0,1,...,d-1$. As
stated in \cite{Collins02a}, their formulation is a \emph{deterministic}
version of hidden variable theory - as was the original treatment by Bell 
\cite{Bell65a} (see Eq. (14) therein), where the hidden variables determine
the \emph{actual} outcomes for measurements. Note that the $C(j,k,l,m)$ are
not required to satisfy\textbf{\ }\emph{locality}\textbf{. }Collins et al 
\cite{Collins02a} point out that their treatment can also be presented in a
more general \emph{non-deterministic} version of HVT, where the hidden
variables merely determine the \emph{probabilities} for measurement outcomes
- such as presented\ in recent work in Refs. \cite{Wiseman07a}, \cite%
{Brunner14a}. The non-deterministic expression for $%
P(j,k,l,m|A_{1},A_{2},B_{1},B_{2})$ is given in Section \ref{SubSection -
NonLocal HVT}. Both approaches lead to the same CGLMP\ inequalities, and the
choice between them is not relevant to the other issues raised in this
paper. \medskip

The fundamental Collins et al \cite{Collins02a} probability $C(j,k,l,m)$ is
based on the simultaneous measurement of two observables for each sub-system
- which is allowed in a classical theory such as hidden variable theory. Due
to the Heisenberg uncertainty principle, such probabilities do not occur in
quantum theory unless the quantum operators for each sub-system commute - so
in general there is no quantum expression for $%
P(j,k,l,m|A_{1},A_{2},B_{1},B_{2})$. Thus in the Collins et al \cite%
{Collins02a} approach, the fundamental probability in the classical theory
which is intended to underlie quantum theory does not itself have a quantum
counterpart. Such an approach is perfectly valid, but it will mean that only
probabilities such as the \emph{marginal probabilities} associated with
measurements of \emph{one} observable per sub-system $P(\alpha _{a},\beta
_{b}|A_{a},B_{b})$ (where $a,b=1,2$ and where $\alpha _{1}\equiv j,\alpha
_{2}\equiv k,\beta _{1}\equiv l,\beta _{2}\equiv m$) could have a quantum
counterpart. The probability $P(j,l|A_{1},B_{1})$ of outcomes $j,l$ for
measurements of observables $A_{1}$, $B_{1}$ (which are for different
sub-systems) is an example. \medskip

In the Collins et al \cite{Collins02a} classical theory such probabilities
are derivable from the $C(j,k,l,m)$ and can be interpreted in terms of
classical measurements. The four different joint probabilities for
measurement outcomes for \emph{one} observable from \emph{each} of the two
sub-systems discussed in Collins et al \cite{Collins02a} are 
\begin{eqnarray}
P(A_{1}
&=&j,B_{1}=l)=P(j,l|A_{1},B_{1})=\tsum%
\limits_{k,m}P(j,k,l,m|A_{1},A_{2},B_{1},B_{2})  \nonumber \\
&=&\tsum\limits_{k,m}C(j,k,l,m)  \nonumber \\
&&...  \nonumber \\
P(A_{2}
&=&k,B_{2}=m)=P(k,m|A_{2},B_{2})=\tsum%
\limits_{j,l}P(j,k,l,m|A_{1},A_{2},B_{1},B_{2})  \nonumber \\
&=&\tsum\limits_{j,l}C(j,k,l,m)  \label{Eq.CollinsTwoObservProb}
\end{eqnarray}%
Here $P(A_{1}=j,B_{1}=l)=P(j,l|A_{1},B_{1})$ is the probability for outcomes 
$j,l$ for measurement of observables $A_{1},B_{1}$ irrespective of the
outcomes for measurement of observables $A_{2},B_{2}$. Such joint
probabilities for one observable for each sub-system are obviously allowed
in the CGLMP\ classical hidden variable theory, given that simultaneous
measurements all four observables $A_{1},A_{2},B_{1},B_{2}$ are allowed.
They are also allowed in quantum theory, since the observables for different
sub-systems are represented by commuting operators. \medskip

One possible classical measurement process is for all specific outcomes $j,l$
for measurement of observables $A_{1},B_{1}$ and the outcomes $k,m$ for
measurements on $A_{2},B_{2}$ to be recorded. The probability for the
outcomes $j,l$ for measurements of $A_{1},B_{1}$ \emph{irrespective} of the
outcomes $k,m$ for measurements of $A_{2},B_{2}$ is then obtained by
dividing the number of results with the same $j,l$ by the total number of
results. Note that as classical measurements can be made without disturbing
the system, the order in which the pairs of measurements for observables $%
A_{1},B_{1}$ and $A_{2},B_{2}$ occur is irrelevant, so doing the
measurements in a different order would be another measurement process that
is classically equivalent for determining $P(A_{1}=j,B_{1}=l)$. Similar
remarks apply to the other three joint probabilities $P(A_{1}=j,B_{2}=m)$, $%
P(A_{2}=k,B_{1}=l)$ and $P(A_{2}=k,B_{2}=m)$. However, there is further way
to measure probabilities such as $P(A_{1}=j,B_{1}=l)$. All specific outcomes 
$j,l$ for measurement of observables $A_{1},B_{1}$ could be recorded but 
\emph{no} measurements would be made on $A_{2},B_{2}$, so again the outcomes 
$k,m$ would be \emph{unrecorded}. The probability for the outcomes $j,l$ for
measurements of $A_{1},B_{1}$ is obtained by dividing the number of results
with the same $j,l$ by the total number of results. Exactly the same
expression as in Eq. (\ref{Eq.CollinsTwoObservProb}) would apply for this
different classical measurement process as for the ones described in the
previous paragraph. Similar remarks apply to the other three joint
probabilities $P(A_{1}=j,B_{2}=m)$, $P(A_{2}=k,B_{1}=l)$ and $%
P(A_{2}=k,B_{2}=m)$. \medskip

Collins et al also introduce probabilities for when the outcomes for
observables of the two sub-systems are either the same or differ by a fixed
amount. The general notation is $P(B_{b}=A_{a}+p(\func{mod}d))$ for the case
where the outcome for $B_{b}$ differs from that for $A_{a}$ by $p(\func{mod}%
d)$. Thus in terms of both the standard notation and in terms of the $%
C(j,k,l,m)$ we have for example 
\begin{eqnarray}
P(A_{1} &=&B_{1})=\sum_{j}P(A_{1}=j,B_{1}=j)  \nonumber \\
&=&\tsum\limits_{j,k,m}C(j,k,j,m)  \label{Eq.ProbTwoOutcomesSame}
\end{eqnarray}%
and 
\begin{eqnarray}
P(B_{1} &=&A_{2}+1)=\sum_{k}P(A_{2}=k,B_{1}=\overline{k+1}(\func{mod}d)) 
\nonumber \\
&=&\tsum\limits_{j,k,m}C(j,k,\overline{k+1}(\func{mod}d),m)
\label{Eq.ProbTwoOutcomesDifferByOne}
\end{eqnarray}%
Here $P(A_{1}=B_{1})=\sum_{j}P(j,j|A_{1},B_{1})$ is the probability that the
outcomes for measurements of $A_{1}$ and $B_{1}$ are the same, irrespective
of what the outcome $j$ is and irrespective of what the outcomes are for
measurements of $A_{2}$ and $B_{2}$. Such probabilities for one observable
for each sub-system are allowed in classical hidden variable theory, since
simultaneous measurements all four observables $A_{1},A_{2},B_{1},B_{2}$ are
allowed, and then all the specific outcomes $j,j$ for measurement of
observables $A_{1},B_{1}$ which are the same and irrespective of the
outcomes $k,m$ for measurements on $A_{2},B_{2}$ can be recorded. Again, the
same expressions would apply if the outcomes for the observables $%
A_{2},B_{2} $ were just left unmeasured or if the order in which the pairs $%
A_{1},B_{1}$ and $A_{2},B_{2}$ were measured was reversed. Similar
considerations apply to $P(B_{1}=A_{2}+1)$ except here the outcomes for
measurements of $A_{2}$ and $B_{1}$ are $k$ and $k+1(\func{mod}d)$,
irrespective of what the outcome $k$ is. \medskip

\section{Local and Non-Local Hidden Variable Theories and CGLMP Inequalities}

\label{Section - Local HVT and CGLMP Inequalities}

In this section the issue of whether the Collins et al \cite{Collins02a}
probabilities are consistent with a \emph{local} hidden variable theory is
considered. We will also show that they are consistent with a very general 
\emph{non-local} hidden variable theory as well\textbf{. }We consider
measurements involving just one observable at a time for both sub-systems,
since this type of measurement is ultimately involved in the CGLMP
inequalities. The application of Fine's theorem \cite{Fine82a} shows that
the probabilities for such measurement outcomes can be expressed in a local
hidden variable theory form. The derivation of a key CGLMP\ inequality then
follows, with the proof not requiring the fundamental probabilities $%
C(j,k,l,m)$ to be determined in accord with\emph{\ local} HVT. We therefore
conclude that the Collins et al \cite{Collins02a} probabilities are
consistent with a \emph{local} hidden variable theory. The presentation set
out below is in terms of the more general non-deterministic HVT, but the
deterministic version can be obtained by replacing the hidden variable
probability distribution $P(\lambda )$ by a delta function. \medskip

\subsection{Consequences of Fine's Theorem}

We can now apply Fine's theorem \ \cite{Fine82a} to express the two
sub-system \ measurement probabilities $P(\alpha _{a},\beta
_{b}|A_{a},B_{b}) $ in a \emph{local hidden variable theory} form. This
result \emph{only} requires the $P(\alpha _{a},\beta _{b}|A_{a},B_{b})$
being given as the \emph{marginal probabilities} based on the four
observable probabilities $C(j,k,l,m)$, as set out in Eq. (\ref%
{Eq.CollinsTwoObservProb}). It is \emph{not required} that the $C(j,k,l,m)$
themselves are given by a \emph{local} hidden variable theory. We thus have 
\begin{equation}
P(\alpha _{a},\beta _{b}|A_{a},B_{b})=\dsum\limits_{\lambda }P(\lambda
)P(\alpha _{a}|A_{a},\lambda )P(\beta _{b}|B_{b},\lambda )
\label{Eq.LocalHVTProbOneObservSubSys}
\end{equation}%
where $P(\lambda )$ is the probability distribution for the hidden variables 
$\lambda $ and $P(\alpha _{a}|A_{a},\lambda )$ and $P(\beta
_{b}|B_{b},\lambda )$ are the separate sub-system probabilities that
measurements of $A_{a}$ and $B_{b}$ lead to outcomes $\alpha _{a}$ and $%
\beta _{b}$ (respectively), when the hidden variable is $\lambda $. The same
hidden variables determine the probabilities for all four cases $%
A_{1},A_{2},B_{1},B_{2}$ separately. Thus in a local HVT the probability $%
P(\alpha _{a},\beta _{b}|A_{a},B_{b},\lambda )$ for the joint measurement
outcome for $A_{a}$ and $B_{b}$ for each hidden variable situation
factorizes into separate probabilities $P(\alpha _{a}|A_{a},\lambda )$ and $%
P(\beta _{b}|B_{b},\lambda )$ for each sub-system. In a non-local hidden
variable theory $P(\alpha _{a},\beta _{b}|A_{a},B_{b},\lambda )$ is not
factorisable. This criterion for locality has been set out in numerous
papers (see for example, \cite{Wiseman07a}, \cite{Reid09a}, \cite{Brunner14a}%
, \cite{Jevtic15a}, \cite{Dalton16a}). Note that this expression shows that
the measurement events where $A_{a}$ leads to outcome $\alpha _{a}$ \ and $%
B_{b}$ leads to outcome $\beta _{b}$ are \emph{classically} correlated.
Expression (\ref{Eq.LocalHVTProbOneObservSubSys}) leads to the well-known 
\emph{no-signalling} result, where the probability of measuring $A_{a}$ to
have an outcome $\alpha _{a}$ when the other sub-system observable $B_{b}$
is either not measured at all or its measurement outcomes $\beta _{b}$ are
left unrecorded, is \emph{independent} of the choice of observable $B_{b}$.

Hence, we see that at the level of the two sub-system measurement
probabilities $P(\alpha _{a},\beta _{b}|A_{a},B_{b})$ for compatible
measurements, which are the basic probabilities that are involved in the
CGLMP Bell inequalities, a local hidden variable theory applies. \medskip

\subsection{Non-Local Hidden Variable Theory}

\label{SubSection - NonLocal HVT}

The basic Collins et al \cite{Collins02a} HVT probabilities $C(j,k,l,m)=$\ $%
P(j,k,l,m|A_{1},A_{2},B_{1},B_{2})$\ for measurements of all four
observables can also be derived from a\textbf{\ }\emph{non-local}\textbf{\ }%
probabilistic HVT. With a distribution $\rho (\lambda )$ of hidden variables 
$\lambda $ determined in the system preparation process with $%
\dsum\limits_{\lambda }\rho (\lambda )=1$, there could be a conditional
probability $P(j,k,l,m|A_{1},A_{2},B_{1},B_{2},\lambda )$ that specifies the
probability that a measurement of all four observables $%
A_{1},A_{2},B_{1},B_{2}$ leads to the outcomes $j,k,l,m$ if the hidden
variable is $\lambda $. In this situation we have 
\begin{equation}
P(j,k,l,m|A_{1},A_{2},B_{1},B_{2})=\dsum\limits_{\lambda
}P(j,k,l,m|A_{1},A_{2},B_{1},B_{2},\lambda )\,\rho (\lambda )
\label{Eq.NonLocalBasicProb}
\end{equation}%
Here it is assumed that the experimenter is able to measure all four
observables in a classical measurement. In a local version of the sub-system
observable measurements we would have 
\begin{equation}
P(j,k,l,m|A_{1},A_{2},B_{1},B_{2},\lambda )=P_{A}(j,k|A_{1},A_{2},\lambda
)P_{B}(l,m|B_{1},B_{2},\lambda )  \label{Eq.LocalProb}
\end{equation}%
\medskip

\subsection{Basic Collins et al Inequality}

\label{SubSection - Collins Inequality}

In this section we will derive a typical Collins et al \cite{Collins02a}
inequality $I\leq 3$ (see Eq. (3) therein). We show that the proof of the
CGLMP inequalities does not depend on the HVT probabilities $C(j,k,l,m)$
themselves satisfying the locality requirement. \medskip

The quantity $I$ is defined by 
\begin{equation}
I=P(A_{1}=B_{1})+P(B_{1}=A_{2}+1)+P(A_{2}=B_{2})+P(B_{2}=A_{1})
\label{Eq.CGLMPQuantity}
\end{equation}%
so \emph{without} invoking a local hidden variable theory for the $%
C(j,k,l,m) $ themselves, we have 
\begin{eqnarray}
I &=&\tsum\limits_{j,k,m}C(j,k,j,m)+\tsum\limits_{j,k,m}C(j,k,\overline{k+1}(%
\func{mod}d),m)+\tsum\limits_{j,k,l}C(j,k,l,k)+\tsum\limits_{j,k,l}C(j,k,l,j)
\nonumber \\
&=&\tsum\limits_{j,k,l,m}C(j,k,l,m)\;\left[ \delta _{l,j}+\delta _{l,k+1(%
\func{mod}d)}+\delta _{m,k}+\delta _{m,j}\right]
\label{Eq.CGLMPQuantityFnOfHVTProbs}
\end{eqnarray}%
Now the quantity in the brackets $\left[ {}\right] $ is never negative and
could only have possible values of $0,1,2,3,4$ in view of the Kronecker
delta only having values of $0$ or $1$. The value $4$ is impossible since
this would require $j=k=m=l$ for the first, third and fourth Kronecker $%
\delta $ to equal $1$ but requires $l=k+1(\func{mod}d)$ for the second
Kronecker $\delta $ to also equal $1$. Hence $\left[ {}\right] \leq 3$ and
thus 
\begin{eqnarray*}
I &\leq &\tsum\limits_{j,k,l,m}C(j,k,l,m)\;[3] \\
&\leq &3
\end{eqnarray*}%
This inequality is valid irrespective of whether or not $C(j,k,l,m)$ itself
is given by a hidden variable theory expression, such as a non-local form or
by a local hidden variable theory expression such as $\dsum\limits_{\lambda
}P(\lambda )P(j,k|A_{1},A_{2},\lambda )P(l,m|B_{1},B_{2},\lambda )$. It is
also valid when the hidden variable probability $P(\lambda )$ is replaced by
a delta function, which turns a non-deterministic HVT into a deterministic
HVT. As the CGLMP inequality has been derived using a HVT which has been
shown to be \emph{local} via Fine's theorem, we conclude that a quantum
state for which $I>3$ would then \emph{violate} Bell locality. \medskip

Collins et al \cite{Collins02a} state that the inequality $I\leq 4$ applies
to general non-local theory. In the case where no hidden variable theory
applies, the quantities $P(A_{1}=B_{1})$, $P(B_{1}=A_{2}+1)$ etc are no
longer related to a set of probabilities $C(j,k,l,m)$, and hence (\ref%
{Eq.CGLMPQuantityFnOfHVTProbs}) no longer applies. In this case each term $%
P(A_{1}=B_{1}),..,P(B_{2}=A_{1})$ could be any positive number between $0$
and $1$, so all that can be concluded is that $I\leq 4$.

However, the inequality can also be obtained from the non-local expression
for $C(j,k,l,m)$. Using Eq. (\ref{Eq.NonLocalBasicProb}) we have instead of (%
\ref{Eq.CGLMPQuantityFnOfHVTProbs})%
\begin{equation}
I=\tsum\limits_{\lambda
}\tsum\limits_{j,k,l,m}P(j,k,l,m|A_{1},A_{2},B_{1},B_{2},\lambda )\,\rho
(\lambda )\;\left[ \delta _{l,j}+\delta _{l,k+1(\func{mod}d)}+\delta
_{m,k}+\delta _{m,j}\right]  \label{Eq.NonLocalCGLMP}
\end{equation}%
and the same steps as before show that $I\leq 3$. \ This also holds if the
local form for $P(j,k,l,m|A_{1},A_{2},B_{1},B_{2},\lambda )$ applies. Thus
the CGLMP Bell inequality can be derived for a very general hidden variable
theory, the only requirement is the existence of a general hidden variable
measurement probability for the joint measurement of all four observables
that depends on the hidden variable. So the violation of the CGLMP Bell
inequality for a particular preparation process leads to the conclusion that
such classical probabilities $P(j,k,l,m|A_{1},A_{2},B_{1},B_{2},\lambda )$
or $C(j,k,l,m)$\ do not exist in regard to describing this process. If the
value for $I$ can be accounted for vis quantum theory, the natural
interpretation would be that quantum theory cannot be under-pinned by a
hidden variable theory. However, several authors including \cite%
{Khrennikov08a}, \cite{Nieuwienhuizen11a} object because other conclusions
are possible. They begin by pointing out that the above probability
description does not satisfy a certain fundamental feature of probability
theory known as\emph{\ contextuality}\textbf{. }This issue will be discussed
in Section \ref{Section - Contextuality Loophole}. \medskip

\section{Possible Quantum Theory Measurement Processes in CLGMP Formalism}

\label{Section - CGLMP and Measurements}

We now examine three possible quantum measurement processes that replicate
classical measurement processes in the Collins et al \cite{Collins02a}
approach for the specific case of $P(A_{1}=j,B_{1}=l)$. These differ by the
order in which measurements on the recorded observables $A_{1}$, $B_{1}$ and
the unrecorded observables $A_{2}$, $B_{2}$ occur, and on whether the
unrecorded observables $A_{2}$, $B_{2}$ are measured at all. For simplicity,
we will assume that both pairs $A_{1}$, $A_{2}$ and $B_{1}$, $B_{2}$ are
incompatible and we note that this applies to the observables in Eq. (13) of
Ref. \cite{Collins02a} which are used in finding a violation of the
inequality $I\leq 3$. Details are set out in Appendix \ref{Appendix -
Quantum Measurement Process - P(A1,B1)}. Treatment of the other
probabilities in Eq.(\ref{Eq.CollinsTwoObservProb}) would be similar. It is
found that for each of these three equivalent classical measurement
processes there is a measurement process in accord with quantum theory that
replicates the measurement process that underpins the CGLMP\ inequalities.
However, \emph{different} quantum theory expressions apply in each case.
Although the three measurement processes are different, it is concluded that
for showing that the Collins et al HVT \cite{Collins02a} does \emph{not}
predict the same results as quantum theory, it is \emph{sufficient} to
demonstrate a CGLMP\ violation for \emph{any one} of the three (or more)
quantum expressions that could be considered. As we will see, the
measurement process in which the pair of observables with \emph{unrecorded
outcomes} are \emph{not} measured at all, leads to quantum expressions for
the probabilities in the CGLMP inequalities that enabled Collins et al \cite%
{Collins02a} to identify a quantum state that violates a CGLMP Bell
inequality.\medskip

\subsection{Quantum Theory Measurements: Two Observables per Sub-System}

\label{SubSection - Q Thy Measts: Two Observ per SubSystem}

We first point out that in the case where the fundamental HVT\ probability $%
P(\alpha ,\beta |A,B,\lambda )$ involves measurements in which there is only
a \emph{single} observable for each sub-system, there is no ambiguity in
relating the classical HVT measurement for $P(\alpha ,\beta |A,B)$ to the
quantum measurement process. In the classical HVT the order of measuring $%
A,B $ is irrelevant, and the same is true in quantum theory. The probability 
$P(\alpha ,\beta |A,B)$ is given in quantum theory by 
\begin{eqnarray}
P_{Q}(\alpha ,\beta |A,B) &=&Tr(\widehat{\Pi }_{\alpha }^{A}\otimes \widehat{%
\Pi }_{\beta }^{B})\widehat{\rho }  \label{Eq.QThyOverrProbBothMeasts} \\
P_{LHVT}(\alpha ,\beta |A,B) &=&\dsum\limits_{\lambda }P(\lambda )P(\alpha
|A,\lambda )P(\beta |B,\lambda )  \label{Eq.LHVTOverrProbBothMeasts}
\end{eqnarray}%
where the projectors $\widehat{\Pi }_{\alpha }^{A}$ and $\widehat{\Pi }%
_{\beta }^{B}$ are the usual quantum projectors onto eigenstates of $A$, $B$
with eigenvalues $\alpha $, $\beta $ respectively, with $\sum_{x}$.$\widehat{%
\Pi }_{x}^{C}=\widehat{1}_{C}$. The LHVT expression is also given for
comparison. The quantum probability is the same irrespective of whether $A$
is measured first and $B$ second, or vice versa. \medskip

We now examine three possible quantum measurement processes that attempt to
replicate classical measurement processes in the Collins et al \cite%
{Collins02a} approach for the specific case of $P(A_{1}=j,B_{1}=l)$ in the
situation dealt with in Collins et al \cite{Collins02a} where there are\emph{%
\ two observables} for each sub-system that have to be considered. \medskip

The \emph{first} possibility would be if measurements on $A_{1},B_{1}$
resulting in outcomes $j,l$ were performed, but measurements on $A_{2},B_{2}$
were $\emph{never}$ performed at all. In this case the joint probability for
the outcomes of measurements on $A_{1},B_{1}$ is given by%
\begin{equation}
P_{Q}(j,l|A_{1},B_{1})=Tr\left( \widehat{\Pi }_{j}^{A_{1}}\otimes \widehat{%
\Pi }_{l}^{B_{1}}\right) \,\widehat{\rho }
\end{equation}%
The notation $P_{Q}(j,l|A_{1},B_{1})$ indicates that \emph{only} $%
A_{1},B_{1} $ measurements were carried out. Note the absence of projection
operators for $A_{2},B_{2}$. This is the same result as (\ref%
{Eq.QThyOverrProbBothMeasts}), where only one observable per sub-system was
ever involved - so $A_{2},B_{2}$ do not exist. Similar considerations apply
to the other probabilities $P(A_{1}=j,B_{2}=m)$, $P(A_{2}=k,B_{1}=l)$ and $%
P(A_{2}=k,B_{2}=m)$. \medskip

The \emph{second} possibility would be if the measurements on $A_{2},B_{2}$
which led to outcomes $k,m$ were performed \emph{first}. The outcomes $k,m$
were then left \emph{unrecorded} and the measurements on $A_{1},B_{1}$ which
led to outcomes $j,l$ were performed \emph{second. }In this case the joint
probability for the outcomes of measurements on $A_{1},B_{1}$ is given by 
\begin{equation}
P_{Q}(j,l,(k,m)|A_{1},B_{1},(A_{2},B_{2})_{1})=\sum_{k,m}Tr\left( \widehat{%
\Pi }_{k}^{A_{2}}\widehat{\Pi }_{j}^{A_{1}}\widehat{\Pi }_{k}^{A_{2}}\otimes 
\widehat{\Pi }_{m}^{B_{2}}\widehat{\Pi }_{l}^{B_{1}}\widehat{\Pi }%
_{m}^{B_{2}}\right) \,\widehat{\rho }  \label{Eq.QThyProbA2B2First}
\end{equation}%
The notation $P_{Q}(j,l,(k,m)|A_{1},B_{1},(A_{2},B_{2})_{1})$ indicates that
the $A_{2},B_{2}$ measurements were carried out $\emph{first}$ and the
results left unrecorded. \medskip

The \emph{third} possibility would be if the measurements on $A_{1},B_{1}$
resulting in outcomes $j,l$ were performed \emph{first}. The measurement of $%
A_{2},B_{2}$ leading to outcomes $k,m$ (which are then left unrecorded) were
performed \emph{second. }In this case the joint probability for the outcomes
of measurements on $A_{1},B_{1}$ is given by 
\begin{equation}
P_{Q}(j,l,(k,m)|A_{1},B_{1},(A_{2},B_{2})_{2})=\sum_{k,m}Tr\left( \widehat{%
\Pi }_{j}^{A_{1}}\widehat{\Pi }_{k}^{A_{2}}\widehat{\Pi }_{j}^{A_{1}}\otimes 
\widehat{\Pi }_{l}^{B_{1}}\widehat{\Pi }_{m}^{B_{2}}\widehat{\Pi }%
_{l}^{B_{1}}\right) \,\widehat{\rho }  \label{Eq.QThyProbA1B1First}
\end{equation}%
The notation $P_{Q}(j,l,(k,m)|A_{1},B_{1},(A_{2},B_{2})_{2})$ indicates that
the $A_{2},B_{2}$ measurements were carried out \emph{second} and the
results left unrecorded. Note the different order of the projection
operators in (\ref{Eq.QThyProbA2B2First}) and (\ref{Eq.QThyProbA1B1First}).
\medskip\ 

The probabilities $P_{Q}(j,l|A_{1},B_{1})$, $%
P_{Q}(j,l,(k,m)|A_{1},B_{1},(A_{2},B_{2})_{1})$ or

$P_{Q}(j,l,(k,m)|A_{1},B_{1},(A_{2},B_{2})_{2})$ for these quantum
measurement processes for $A_{1},B_{1}$ to have outcomes $j,l$ are \emph{not}
the same. The same applies to the final density operators (see Appendix \ref%
{Appendix - Quantum Measurement Process - P(A1,B1)}). This confirms that the
three classically equivalent measurement processes that equally determine
the probability $P(A_{1}=j,B_{1}=l)$.can each be described via quantum
theory, but the three quantum theory predictions are different. There are
even further possibilities that could have been considered, such as
involving measuring observables $A_{1},B_{2}$ resulting in outcomes $j,m$
are measured \emph{first} with the outcome $m$ left unrecorded, followed by
measurement of $A_{2},B_{1}$ leading to outcomes $k,l$ with the outcome $k$
then left unrecorded. So which one is to be chosen to give the quantum
theory analogue of the Collins et al \cite{Collins02a} quantity $%
P(A_{1}=j,B_{1}=l)$ ? However, as pointed out previously, it is \emph{%
sufficient} to show that a CGLMP\ inequality is violated for any \emph{one }%
of the possible\emph{\ quantum measurement processes} and for any \emph{one
quantum state} to demonstrate that the Collins et al HVT \cite{Collins02a}
cannot predict the same results as quantum theory.\medskip

Collins et al \cite{Collins02a} actually make the comparison of the CGLMP
hidden variable theory predictions with those from quantum theory by
choosing the measurement processes to be those where the unrecorded pair of
observables are just \emph{not measured} at all - the \emph{first }(and
simplest) possibility discussed above. It can easily be \emph{confirmed}
from the quantum theory probability expression set out in Eq. (14) therein
for $P_{QM}(A_{a}=k,B_{b}=l)$ that this is the approach that Collins et al
adopted. Hence Collins et al \cite{Collins02a} compare the LHVT expression $%
P(A_{1}=j,B_{1}=l)$ with the quantum theory expression for $%
P_{Q}(A_{1}=j,B_{1}=l)$ 
\begin{equation}
P(A_{1}=j,B_{1}=l)=\tsum\limits_{k,m}C(j,k,l,m)\equiv Tr\left( \widehat{\Pi }%
_{j}^{A_{1}}\otimes \widehat{\Pi }_{l}^{B_{1}}\right) \,\widehat{\rho }%
=P_{Q}(j,l|A_{1},B_{1})  \label{Eq.HVTQuantumTheoryEquivalences1}
\end{equation}%
\medskip

For the final probabilities $P(A_{1}=B_{1})$, $P(B_{1}=A_{2}+1)$ etc that
appear in the CGLMP inequalities, similar considerations apply and quantum
theory expressions such as 
\begin{eqnarray}
P(A_{1} &=&B_{1})=\tsum\limits_{j,k,m}C(j,k,j,m)\equiv Tr\sum_{j}\left( 
\widehat{\Pi }_{j}^{A_{1}}\otimes \widehat{\Pi }_{j}^{B_{1}}\right) \,%
\widehat{\rho }  \nonumber \\
&=&\sum_{j}P_{Q}(j,j|A_{1},B_{1})  \nonumber \\
P(B_{1} &=&A_{2}+1)=\tsum\limits_{j,k,m}C(j,k,\overline{k+1}(\func{mod}d),m)
\nonumber \\
&\equiv &Tr\sum_{k}\left( \widehat{\Pi }_{k}^{A_{2}}\otimes \widehat{\Pi }_{%
\overline{k+1}(\func{mod}d)}^{B_{1}}\right) \,\widehat{\rho }  \nonumber \\
&=&\sum_{k}P_{Q}(k,\overline{k+1}(\func{mod}d)|A_{2},B_{1})  \nonumber \\
&&.  \label{Eq.HVTQuantumTheoryEquivalences2}
\end{eqnarray}%
have been assumed in Collins et al \cite{Collins02a}. \medskip

Since the Bell inequalities involve \emph{joint mean values} derived from
the \emph{joint probabilities} such as $P(\alpha ,\beta |A,B)$ given by 
\begin{equation}
\left\langle A\otimes B\right\rangle =\tsum\limits_{\alpha ,\beta }(\alpha
\times \beta )\;P(\alpha ,\beta |A,B)  \label{Eq.JointMean}
\end{equation}%
then the LHVT expression based on (\ref{Eq.LHVTOverrProbBothMeasts}) and the
quantum theory expression based on (\ref{Eq.QThyOverrProbBothMeasts}) can be
compared in terms of the same measurement process. The mean value
expressions for LHVT and quantum theory are 
\begin{eqnarray}
\left\langle A\otimes B\right\rangle _{LHVT} &=&\dsum\limits_{\lambda
}P(\lambda )\tsum\limits_{\alpha ,\beta }(\alpha \times \beta )\;P(\alpha
|A,\lambda )\;P(\beta |B,\lambda )  \nonumber \\
&=&\dsum\limits_{\lambda }P(\lambda )\left\langle (A)_{\lambda
}\right\rangle _{LHVT}\left\langle (B)_{\lambda }\right\rangle _{LHVT} 
\nonumber \\
\left\langle A\otimes B\right\rangle _{Q} &=&Tr\tsum\limits_{\alpha ,\beta
}(\alpha \times \beta )\;(\widehat{\Pi }_{\alpha }^{A}\otimes \widehat{\Pi }%
_{\beta }^{B})\widehat{\rho }=Tr(\widehat{A}\otimes \widehat{B})\widehat{%
\rho }  \label{Eq.JointMeanB}
\end{eqnarray}%
in an obvious notation.\medskip

Thus, in spite of the CGLMP Bell inequalities being based on expressions
such as $P(A_{1}=B_{1})$, $P(B_{1}=A_{2}+1)$ which have several possible
equivalents in quantum theory, conclusions that certain quantum states and
related observables lead to violations of Bell locality can still be
made.\medskip

\subsection{Quantum State Violating Collins Inequality}

An example of a quantum state (see their Eq. 12)) that violates the
inequality $I\leq 3$ is considered by Collins et al \cite{Collins02a}, based
on quantum expressions in Eq. (\ref{Eq.HVTQuantumTheoryEquivalences2}). This
is the maximally entangled state for two $d$ dimensional sub-systems. Such a
case could apply to two sub-systems with the same spin $s$, for which the
spin eigenstates are $\left\vert s,m\right\rangle $, where $m=-s,..,+s$. For
the (un-normalized) state $\dsum\limits_{m=-s}^{s}\left\vert
s,m\right\rangle _{A}\left\vert s,m\right\rangle _{B}$ the quantum
expression for $I$ is found to be greater than $3$ for all $d=2s+1$. In the
case where $s$ is large, this corresponds to a Bell inequality violation in
a \emph{macroscopic} system - a result of some significance, as Bell
non-locality is normally confined to microscopic systems. \medskip

However, this violation involved introducing physical quantities $%
A_{1},A_{2},B_{1},B_{2}$ as Hermitian operators defined by their eigenvalues
and eigenvectors (see Eq. (13) in Ref \cite{Collins02a}), the latter being
linear combinations of the $\left\vert s,m\right\rangle _{A(B)}$ .However,
as the operators turn out to be \emph{off-diagonal} in these basis states,
it is \emph{not} obvious what physical observable they correspond to.
\medskip

No experimental tests of the CGLMP\ Bell inequalities have been carried out.
However, it may be possible to test the CGLMP inequalities in Bose-Einstein
condensates in cold atomic gases based on double-well potentials, with two
localized modes per well associated with different hyperfine states and with
the $\left\vert s,m\right\rangle _{A}$ and $\left\vert s,m\right\rangle _{B}$
being Schwinger spin states for each the sub-system. In principle $s$ can be
large when $N$ is large. The two different observables for each sub-system
could be defined in terms of the Schwinger spin states.\medskip

\section{The Contextuality Loophole}

\label{Section - Contextuality Loophole}

The violation of the CGLMP Bell inequality such as $I\leq 3$\ for some
physical states and choices of measurement processes shows that a HVT\
probability such as $C(j,k,l,m)$ or $P(j,k,l,m|A_{1},A_{2},B_{1},B_{2},%
\lambda )$ that \ describes the outcome of measuring the four observables $%
A_{1},A_{2},B_{1},B_{2}$ does not exist, at least for such states and
processes, and therefore such a HVT approach based on such a single
probability function that describes all the separate measurements has no
general validity. This follows because (as we have shown in Section \ref%
{SubSection - Collins Inequality}) the existence of such a single
probability function results in the CGLMP\ inequality. If a quantum theory
treatment can account for the measurement results for all the physical
states, the widely accepted conclusion is that only quantum theory provides
the correct approach and it is not underpinned by HVT, local or non-local.
However, a number of authors (such as in Refs. \cite{Hess05a}, \cite{Hess09a}%
, \cite{Khrennikov08a}, \cite{Khrennikov15a}, \cite{Nieuwienhuizen11a}, \cite%
{Kupczynski04a}) have objected, on the grounds that this conclusion is not
the only one that could be drawn. \medskip

As has also been pointed out in the present paper, the quantities involved
in Bell inequalities do not arise from a single physical measurement. For
example, there are four different physically allowed measurements involved
in the CGLMP\ inequality $I\leq 3$, namely the measurements of compatible
pairs $(A_{1},B_{1})$ , $(B_{1},A_{2})$ ,$(A_{2},B_{2})$ ,$(B_{2},A_{1})$.
Each pair is said to represent a different experimental\textbf{\ }\emph{%
context}\textbf{. }Hence, according to classical Kolmogorov probability
theory, these are different experiments, and if the measurement outcomes for
each pair is to be interpreted in terms of a hidden variable theory, each
pair should have its own hidden variables and related probability
distribution. Thus we would have (in an obvious notation) 
\begin{eqnarray}
P(j,l|A_{1},B_{1}) &=&\dsum\limits_{\lambda _{1,1}}P(j,l|A_{1},B_{1},\lambda
_{1,1})\rho _{1,1}(\lambda _{1,1})  \nonumber \\
P(k,l|A_{2},B_{1}) &=&\dsum\limits_{\lambda _{2,1}}P(k,l|A_{2},B_{1},\lambda
_{2,1})\rho _{2,1}(\lambda _{2,1})  \nonumber \\
P(k,m|A_{2},B_{2}) &=&\dsum\limits_{\lambda _{2,2}}P(k,m|A_{2},B_{2},\lambda
_{2,2})\rho _{2,2}(\lambda _{2,2})  \nonumber \\
P(j,m|A_{1},B_{2}) &=&\dsum\limits_{\lambda _{1,2}}P(j,m|A_{1},B_{2},\lambda
_{1,2})\rho _{1,2}(\lambda _{1,2})  \label{Eq.Kolmog}
\end{eqnarray}%
With this constraint that the underlying probabilities $\rho _{i,j}(\lambda
_{i,j})$, $P(\alpha _{i},\beta _{j}|A_{i},B_{j},\lambda _{i,j})$ satisfy the
contextuality requirement of being different, the resulting expression for $%
I $\ in the CGLMP\ inequality can no longer be shown to necessarily satisfy
the inequality $I\leq 3$. However it might do so. Obviously there must be an
inequality $I\leq 4$ for any choice if the individual factors in (\ref%
{Eq.Kolmog}) - subject to the usual probability sum rules. An analogous
consideration of the correlators in the CHSH\ Bell inequality is presented
in Ref. \cite{Nieuwienhuizen11a}, where it is pointed out that if separate
probability functions are introduced for the four different settings
involved, then the CHSH inequality no longer necessarily applies, though
again it might.\medskip

From the point of view where the measurement of each pair of observables is
regarded as representing a different experimental context,\ these examples
show that a violation of a Bell inequality such as CGLMP or CHSH\textbf{\ }%
\emph{could}\textbf{\ }still be interpreted as being consistent with the
existence of a HVT interpretation of measurement outcomes, but it would have
to be a HVT formalism that complies with the contextuality requirement in
which each of the several separate experiment contexts in the inequality
involves its own separate HVT probabilities. As far as the author is aware,
no such HVT has yet been constructed. So applying this\textbf{\ }\emph{%
contextuality loophole}\textbf{, }a hidden variable theory could still be a
possible approach to treating measurements in spite of a Bell inequality
violation. Bell inequalities would hence be regarded as rather useless. This
of course is consistent with quantum theory also being a valid approach - a
point not being disputed.\medskip

However, there is another point of view based on the idea that a meaningful
HVT\textbf{\ }\emph{requires a single}\textbf{\ }probability function such
as $C(j,k,l,m)$ or $P(j,k,l,m|A_{1},A_{2},B_{1},B_{2},\lambda )$ that\
describes the outcome of measuring the four classical observables $%
A_{1},A_{2},B_{1},B_{2}$\ irrespective of whether they are all being
measured together or one pair after another. After all, the hidden variables
are intended to determine (at least probabilisticly) the actual values the
observables have, and those values for each observable would not change
between two measurements done at the same time just because some other
observable was being measured or not. One could regard the sequence of
pairwise measurements as one big experiment with its own context. From this
point of view, Bell and others are entitled to frame HVT in terms of a
single probability function, in which case the violation of a Bell
inequality such as CGLMP\textbf{\ }\emph{does}\textbf{\ }show that HVT does
not describe experiment.\textbf{\ }\medskip

In this paper we have discussed the quantum measurement of the quantities
involved in the CGLMP Bell inequality in terms of projective measurements of
pairs of compatible observables, one from each sub-system. This enables
standard quantum theory expressions for the measured outcomes to be used to
determine the predicted results for the left side of the CGLMP inequality
for various quantum states in order to see whether the inequality can be
violated. When this occurs there is no single classical HVT joint
probability that can account for the quantum predictions based on the
different measurement contexts associated with the four different pairs of
sub-system observables. However, within the framework of \textit{generalised
quantum measurement theory} (see Ref. \cite{Barnett09a} for example), it is
possible to describe simultaneous measurements of incompatible observables
quantum mechanically via so-called positive operator value measures (POVMs).
In generalised measurement theory, the measured outcomes are less precise
than in projective measurements. This single context approach has been used
to consider the CHSH Bell inequalities \cite{vanHeusden19a} for two spin $%
1/2 $ systems and the Mermin inequality for three particle GHZ states \cite%
{Son05a}, and as might be expected, it is found that no violations of the
Bell inequalities occur. In terms of this approach, all four terms in the
CGLMP Bell inequality could be measured in a single context via the
measurement process associated with the several POVMs, and presumably no
violation of the CGLMP inequality would occur here as well. However, in the
present paper we consider a HVT based on Einstein's original idea that the
values of the hidden variables are created in the preparation process and
then determine the measured outcomes for all the observables - no matter in
what context or sequence they are measured, and compare this form of HVT\
with quantum theory predictions based on standard projective measurements -
which after all, are the ones where definite outcomes for quantum
measurements occur, just as are required for the classical observables in
Einstein's HVT. \medskip

\section{Conclusion}

\label{Section - Summary}

The significance of the CGLMP (Bell) inequalities as possible tests for
showing that quantum theory is underpinned by a local hidden variable theory
has been discussed. The question of whether Collins et al actually used a
local form of hidden variable theory has been examined, and it is concluded
that although the fundamental hidden variable theory probabilities $%
C(j,k,l,m)$ of measurement outcomes for two observables in each sub-system
are deterministic and may not satisfy the requirements for a local HVT, the
application of Fine's theorem shows that the marginal probabilities $%
P(\alpha _{a},\beta _{b}|A_{a},B_{b})$ for measurements of one observable in
both of the two sub-systems do satisfy the locality requirement. Leaving
aside the contextuality loophole issue, then since the CGLMP inequalities
involve measurements based on the marginal probabilities which satisfy
locality requirements, it is concluded that the CGLMP inequalities do
provide a test for Bell locality applying to pairs of compatible observables
- as the authors stated. The proof of the CGLMP inequalities does not depend
on the HVT probabilities $C(j,k,l,m)$ themselves satisfying the locality
requirement. However, we also showed that a non-local, non-deterministic
interpretation of the HVT\ probabilities $C(j,k,l,m)$\ also leads to the
CGLMP inequalities, so their violation provides a test for a Bell non-local
HVT applying to all observables.

The CGLMP (Bell) inequalities are based on a form of hidden variable theory
(HVT) that allows for simultaneous measurements of pairs of observables that
correspond to non-commuting quantum operators. However, although this is
allowed in a classical probability theory, it does lead to the CGLMP Bell
inequalities being based on expressions for which a number of different
quantum theory expressions apply, corresponding to different measurement
processes that would have been equivalent in classical physics.
Fundamentally, this is because quantum measurements change the quantum state
whereas classical measurements leave the state unchanged, and for
observables whose outcomes are unrecorded whether measurements of these
observables are made and discarded, differs in quantum theory from the case
where the measurements are not made at all. However, conclusions that
certain quantum states and related observables lead to violations of HVT can
still be made based on any one of the possible quantum theory expressions
that replicates an equivalent classical measurement processes that could
determine the probabilities in the CGLMP\ inequalities. The most convenient
quantum measurement process is the one where pairs of observables whose
results are to be left unrecorded are never measured at all.

Based on the expression for this quantum measurement process, Collins et al
have identified a quantum state that violates the CGLMP inequality $I\leq 3$%
. The state involved could apply to macroscopic systems. However, the
observables found by Collins et al to be associated with the CGLMP
inequality $I\leq 3$ violation have no obvious physical interpretation.
Leaving aside the issue\textbf{s} of interpreting the observables and the
contextuality loophole, it is concluded that the CGLMP inequalities have
been shown to rule out local hidden variable theories and even non-local
hidden variable theories as well. This conclusion is based on HVT involving
a single probability function that describes the measurement outcomes of
measurements for all four observables, both all together and in the four
separate measurement contexts where particular pairs of compatable
observables are being measured. However, as discussed in Section \ref%
{Section - Contextuality Loophole}, if the contextuality loophole is
accepted, CGLMP inequality violation does not rule out a hidden variable
theory approach to treating the CGLMP measurements, though separate HVT
probabilities would be required for each of the four pair-based measurement
contexts - as in standard Kolmogorov probability theory, and Bell
inequalities become rather pointless. But on the other hand, as discussed in
Section \ref{Section - Contextuality Loophole}, it is reasonable that a
meaningful HVT should involve only a single probability function able to
describe the outcomes for all observables, irrespective of whether they are
all measured together or in one pair after another, and the sequence of
pairwise measurements can be regarded as one big experiment. Classical
measurements as in HVT are not supposed to disturb the values of the
observables, unlike quantum projective measurements. From this perspective,
the violation of a Bell inequality such as CGLMP would establish that HVT
was not capable of accounting for the experimental results - which is the
widely held viewpoint. We have not examined the CGLMP Bell inequalities from
the point of general quantum measurement theory approach based on POVMs, as
we regard the comparison of HVT should be with quantum theory based on the
more precise standard projective measurements. We also point out that CGLMP\
tests might be carried out in macroscopic systems such as Bose-Einstein
condensates of atomic gases with two hyperfine components in a double
potential well, as states and observables based on the Schwinger spin states
for each well could be suitable. \medskip

\section{Acknowledgements}

The author is grateful for helpful discussions with Drs E Cavalcanti, R Y
Teh and J Vaccaro, and Professors S M Barnett and B M Garraway. Financial
support from University of Glasgow and Imperial College, London during the
conduct of this research is acknowledged. The author is also grateful to two
referees for raising issues that have resulted in a somewhat different
conclusion about CGLMP inequalities, and to one referee for directing his
attention to the contextuality loophole.\textbf{\ }\medskip

\section{Author Contribution Statement}

This paper was entirely written by B J Dalton. 

\section{Appendix - Quantum Measurement Processes Replicating $%
P(A_{1},B_{1}) $ - Four Observables}

\label{Appendix - Quantum Measurement Process - P(A1,B1)}

The \emph{first} possibility is that $A_{2},B_{2}$ are never measured at
all. The probability $P_{Q}(j,l|A_{1},B_{1})$ for measurement of observables 
$A_{1},B_{1}$ \emph{alone} that results in outcomes $j,l$ would be%
\begin{equation}
P_{Q}(j,l|A_{1},B_{1})=Tr\left( \widehat{\Pi }_{j}^{A_{1}}\otimes \widehat{%
\Pi }_{l}^{B_{1}}\right) \,\widehat{\rho }  \label{Eq.ProbA1B1OnlyMeasured}
\end{equation}%
and quantum density operator following this measurement just of observables $%
A_{1},B_{1}$ alone would be 
\begin{equation}
\widehat{\rho }_{j,l}\ =\left( \widehat{\Pi }_{j}^{A_{1}}\otimes \widehat{%
\Pi }_{l}^{B_{1}}\right) \,\widehat{\rho }\,\left( \widehat{\Pi }%
_{j}^{A_{1}}\otimes \widehat{\Pi }_{l}^{B_{1}}\right) /Tr\left( \widehat{\Pi 
}_{j}^{A_{1}}\otimes \widehat{\Pi }_{l}^{B_{1}}\right) \,\widehat{\rho }
\label{Eq.FinalStateA1B1OnlyMeasured}
\end{equation}

The \emph{second} possibility would be if the measurements on $A_{2},B_{2}$
which led to outcomes $k,m$ were performed \emph{first}, the results left
unrecorded and then $A_{1},B_{1}$ measured leading to outcomes $j,l$. In
this case after the measurements on $A_{2},B_{2}$ are carried out with
outcomes $k,m$, the original density operator $\widehat{\rho }$ changes to $%
\widehat{\rho }^{\#}$ where (see Sect 8.3.1 in Ref \cite{Copenhagen}) 
\begin{equation}
\widehat{\rho }^{\#}\ =\left( \widehat{\Pi }_{k}^{A_{2}}\otimes \widehat{\Pi 
}_{m}^{B_{2}}\right) \,\widehat{\rho }\left( \widehat{\Pi }%
_{k}^{A_{2}}\otimes \widehat{\Pi }_{m}^{B_{2}}\right) /Tr\left( \widehat{\Pi 
}_{k}^{A_{2}}\otimes \widehat{\Pi }_{m}^{B_{2}}\right) \,\widehat{\rho }
\end{equation}%
If the results for all the outcomes $k,m$ are then left unrecorded the
density operator changes again to $\widehat{\rho }^{\#\#}$ where%
\begin{equation}
\widehat{\rho }^{\#\#}\ =\sum_{k,m}\widehat{\rho }^{\#}\times Tr\left( 
\widehat{\Pi }_{k}^{A_{2}}\otimes \widehat{\Pi }_{m}^{B_{2}}\right) \,%
\widehat{\rho }=\sum_{k,m}\left( \widehat{\Pi }_{k}^{A_{2}}\otimes \widehat{%
\Pi }_{m}^{B_{2}}\right) \,\widehat{\rho }\left( \widehat{\Pi }%
_{k}^{A_{2}}\otimes \widehat{\Pi }_{m}^{B_{2}}\right)
\end{equation}%
This of course also differs from the original density operator $\widehat{%
\rho }$. So it is not as if $A_{2},B_{2}$ had never been measured at all.
The quantum probability for measurements on $A_{2},B_{2}$ which led to all
outcomes $k,m$ is obviously $P_{Q}((k,m)|A_{2},B_{2})=\sum_{k,m}Tr\left( 
\widehat{\Pi }_{k}^{A_{2}}\otimes \widehat{\Pi }_{m}^{B_{2}}\right) \,%
\widehat{\rho }=1$, \ where the notation $(k,m)$ indicates outcome events
for all $k,m$.

The probability that the subsequent measurement of observables $A_{1},B_{1}$
resulting in outcomes $j,l$ will be given by the conditional probability 
\begin{eqnarray}
Tr\left( \widehat{\Pi }_{j}^{A_{1}}\otimes \widehat{\Pi }_{l}^{B_{1}}\right)
\,\widehat{\rho }^{\#\#} &=&\sum_{k,m}Tr\left( \widehat{\Pi }%
_{j}^{A_{1}}\otimes \widehat{\Pi }_{l}^{B_{1}}\right) \,\left( \widehat{\Pi }%
_{k}^{A_{2}}\otimes \widehat{\Pi }_{m}^{B_{2}}\right) \,\widehat{\rho }%
\left( \widehat{\Pi }_{k}^{A_{2}}\otimes \widehat{\Pi }_{m}^{B_{2}}\right) 
\nonumber \\
&=&\sum_{k,m}Tr\left( \widehat{\Pi }_{k}^{A_{2}}\widehat{\Pi }_{j}^{A_{1}}%
\widehat{\Pi }_{k}^{A_{2}}\otimes \widehat{\Pi }_{m}^{B_{2}}\widehat{\Pi }%
_{l}^{B_{1}}\widehat{\Pi }_{m}^{B_{2}}\right) \,\widehat{\rho }
\end{eqnarray}%
so that the overall quantum probability for the event where measurement of $%
A_{1},B_{1}$ results in outcomes $j,l$ after measurement of $A_{2},B_{2}$
results in all outcomes $k,m$ is obtained by multiplying this conditional
probability by $P_{Q}((l,m)|A_{2},B_{2})=1$, and is given by 
\begin{equation}
P_{Q}(j,l,(k,m)|A_{1},B_{1},(A_{2},B_{2})_{1})=\sum_{k,m}Tr\left( \widehat{%
\Pi }_{k}^{A_{2}}\widehat{\Pi }_{j}^{A_{1}}\widehat{\Pi }_{k}^{A_{2}}\otimes 
\widehat{\Pi }_{m}^{B_{2}}\widehat{\Pi }_{l}^{B_{1}}\widehat{\Pi }%
_{m}^{B_{2}}\right) \,\widehat{\rho }
\label{Eq.ProbMeasureA2B2FirstA1B1Second}
\end{equation}%
The notation $P_{Q}(j,l,(k,m)|A_{1},B_{1},(A_{2},B_{2})_{1})$ indicates that
the $A_{2},B_{2}$ measurements were carried out $\emph{first}$ and the
results left unrecorded. The final measurement of observables $A_{1},B_{1}$
leading to outcomes $j,l$ results in the further change of the density
operator from $\widehat{\rho }^{\#\#}$ to $\widehat{\rho }^{\#\#\#}$ where%
\begin{eqnarray}
\widehat{\rho }^{\#\#\#}\ &=&\sum_{k,m}\left( \widehat{\Pi }_{j}^{A_{1}}%
\widehat{\Pi }_{k}^{A_{2}}\otimes \widehat{\Pi }_{l}^{B_{1}}\widehat{\Pi }%
_{m}^{B_{2}}\right) \,\widehat{\rho }\left( \widehat{\Pi }_{k}^{A_{2}}%
\widehat{\Pi }_{j}^{A_{1}}\otimes \widehat{\Pi }_{m}^{B_{2}}\widehat{\Pi }%
_{l}^{B_{1}}\right) /  \nonumber \\
&&\left\{ \sum_{k,m}Tr\left( \widehat{\Pi }_{k}^{A_{2}}\widehat{\Pi }%
_{j}^{A_{1}}\widehat{\Pi }_{k}^{A_{2}}\otimes \widehat{\Pi }_{m}^{B_{2}}%
\widehat{\Pi }_{l}^{B_{1}}\widehat{\Pi }_{m}^{B_{2}}\right) \,\widehat{\rho }%
\right\}  \label{Eq.FinalStateA2B2FirstA1B1Second}
\end{eqnarray}

Next we consider a \emph{third} possibility for what happens in quantum
theory in a measurement process that replicates the classical measurement
process on which $P(A_{1}=j,B_{1}=l)$ is based when observables $A_{1},B_{1}$
resulting in outcomes $j,l$ are measured \emph{first}, followed by
measurement of $A_{2},B_{2}$ leading to outcomes $l,m$, which are then left
unrecorded.

After the first measurement the original density operator $\widehat{\rho }$
changes to $\widehat{\rho }^{\&}$ where (see Sect 8.3.1 in Ref \cite%
{Copenhagen}) 
\begin{equation}
\widehat{\rho }^{\&}\ =\left( \widehat{\Pi }_{j}^{A_{1}}\otimes \widehat{\Pi 
}_{l}^{B_{1}}\right) \,\widehat{\rho }\left( \widehat{\Pi }%
_{j}^{A_{1}}\otimes \widehat{\Pi }_{l}^{B_{1}}\right) /Tr\left( \widehat{\Pi 
}_{j}^{A_{1}}\otimes \widehat{\Pi }_{l}^{B_{1}}\right) \,\widehat{\rho }
\end{equation}%
The probability for measurement of observables $A_{1},B_{1}$ resulting in
outcomes $j,l$ is given by 
\begin{equation}
P_{Q}(j,l|A_{1},B_{1})=Tr\left( \widehat{\Pi }_{j}^{A_{1}}\otimes \widehat{%
\Pi }_{l}^{B_{1}}\right) \,\widehat{\rho }
\end{equation}%
After the subsequent measurement of $A_{2},B_{2}$ leading to outcomes $k,m$
the density operator becomes 
\begin{eqnarray}
\widehat{\rho }^{\&\&}\ &=&\left( \widehat{\Pi }_{k}^{A_{2}}\widehat{\Pi }%
_{j}^{A_{1}}\otimes \widehat{\Pi }_{m}^{B_{2}}\widehat{\Pi }%
_{l}^{B_{1}}\right) \,\widehat{\rho }\left( \widehat{\Pi }_{j}^{A_{1}}%
\widehat{\Pi }_{k}^{A_{2}}\otimes \widehat{\Pi }_{l}^{B_{1}}\widehat{\Pi }%
_{m}^{B_{2}}\right) /  \nonumber \\
&&\left\{ Tr\left( \widehat{\Pi }_{j}^{A_{1}}\widehat{\Pi }_{k}^{A_{2}}%
\widehat{\Pi }_{j}^{A_{1}}\otimes \widehat{\Pi }_{l}^{B_{1}}\widehat{\Pi }%
_{m}^{B_{2}}\widehat{\Pi }_{l}^{B_{1}}\right) \,\,\widehat{\rho }\right\}
\end{eqnarray}%
and this outcome occurs with a conditional probability $Tr\left( \widehat{%
\Pi }_{k}^{A_{2}}\otimes \widehat{\Pi }_{m}^{B_{2}}\right) \,\widehat{\rho }%
^{\&}$.

The conditional probability for all outcomes $k,m$ for measurements of $%
A_{2},B_{2}$ following measurements of $A_{1},B_{1}$ resulting in outcomes $%
j,l$ will then be%
\begin{eqnarray}
&&\sum_{k,m}Tr\left( \widehat{\Pi }_{k}^{A_{2}}\otimes \widehat{\Pi }%
_{m}^{B_{2}}\right) \,\widehat{\rho }^{\&}  \nonumber \\
&=&\sum_{k,m}Tr\left( \widehat{\Pi }_{k}^{A_{2}}\otimes \widehat{\Pi }%
_{m}^{B_{2}}\right) \,\left( \widehat{\Pi }_{j}^{A_{1}}\otimes \widehat{\Pi }%
_{l}^{B_{1}}\right) \,\widehat{\rho }\,\left( \widehat{\Pi }%
_{j}^{A_{1}}\otimes \widehat{\Pi }_{l}^{B_{1}}\right) /  \nonumber \\
&&Tr\left( \widehat{\Pi }_{j}^{A_{1}}\otimes \widehat{\Pi }%
_{l}^{B_{1}}\right) \,\widehat{\rho }  \nonumber \\
&=&\sum_{k,m}Tr\left( \widehat{\Pi }_{j}^{A_{1}}\widehat{\Pi }_{k}^{A_{2}}%
\widehat{\Pi }_{j}^{A_{1}}\otimes \widehat{\Pi }_{l}^{B_{1}}\widehat{\Pi }%
_{m}^{B_{2}}\widehat{\Pi }_{l}^{B_{1}}\right) \,\,\widehat{\rho }\,/Tr\left( 
\widehat{\Pi }_{j}^{A_{1}}\otimes \widehat{\Pi }_{l}^{B_{1}}\right) \,%
\widehat{\rho }
\end{eqnarray}%
The overall quantum probability for the event where measurement of $%
A_{1},B_{1}$ results in outcomes $j,l$ before measurement of $A_{2},B_{2}$
results in all outcomes $k,m$ is obtained by multiplying this conditional
probability by $P_{Q}(j,l|A_{1},B_{1})$ and is given by 
\begin{equation}
P_{Q}(j,l,(k,m)|A_{1},B_{1},(A_{2},B_{2})_{2})=\sum_{k,m}Tr\left( \widehat{%
\Pi }_{j}^{A_{1}}\widehat{\Pi }_{k}^{A_{2}}\widehat{\Pi }_{j}^{A_{1}}\otimes 
\widehat{\Pi }_{l}^{B_{1}}\widehat{\Pi }_{m}^{B_{2}}\widehat{\Pi }%
_{l}^{B_{1}}\right) \,\,\widehat{\rho }
\end{equation}%
The notation $P_{Q}(j,l,(k,m)|A_{1},B_{1},(A_{2},B_{2})_{2})$ indicates that
the $A_{2},B_{2}$ measurements were carried out second and the results left
unrecorded. Note the different expressions for $%
P_{Q}(j,l,(k,m)|A_{1},B_{1},(A_{2},B_{2})_{2})$ and $%
P_{Q}(j,l,(k,m)|A_{1},B_{1},(A_{2},B_{2})_{1})$, which are both different to 
$P_{Q}(j,l|A_{1},B_{1})$, the probability for just measuring $A_{1},B_{1}$
alone. The density operator after the $A_{2},B_{2}$ measurements were
carried out and the outcomes $l,m$ are left unrecorded changes from $%
\widehat{\rho }^{\&\&}$ to $\widehat{\rho }^{\&\&\&}$ where 
\begin{equation}
\widehat{\rho }^{\&\&\&}\ =\sum_{k,m}\left( \widehat{\Pi }_{k}^{A_{2}}%
\widehat{\Pi }_{j}^{A_{1}}\otimes \widehat{\Pi }_{m}^{B_{2}}\widehat{\Pi }%
_{l}^{B_{1}}\right) \,\widehat{\rho }\left( \widehat{\Pi }_{j}^{A_{1}}%
\widehat{\Pi }_{k}^{A_{2}}\otimes \widehat{\Pi }_{l}^{B_{1}}\widehat{\Pi }%
_{m}^{B_{2}}\right) /\left\{ Tr\left( \widehat{\Pi }_{j}^{A_{1}}\otimes 
\widehat{\Pi }_{l}^{B_{1}}\right) \,\widehat{\rho }\right\}
\end{equation}

In the general case where the two operators of each sub-system do not
commute, the results for $\widehat{\rho }^{\#\#\#}$ and $%
P_{Q}(j,l,(k,m)|A_{1},B_{1},(A_{2},B_{2})_{1})$ or $\widehat{\rho }^{\&\&\&}$
and $P_{Q}(j,l,(k,m)|A_{1},B_{1},(A_{2},B_{2})_{2})$ are not the same as if
measurements on $A_{2},B_{2}$ had \emph{never} taken place at all. .So not
only does the measurement probability $P_{Q}(j,l|A_{1},B_{1})$ differ from $%
P_{Q}(j,l,(k,m)|A_{1},B_{1},(A_{2},B_{2})_{1})$ or $%
P_{Q}(j,l,(k,m)|A_{1},B_{1},(A_{2},B_{2})_{2})$, but the final quantum
states $\widehat{\rho }_{j,l}$ or $\widehat{\rho }^{\#\#\#}$ or $\widehat{%
\rho }^{\&\&\&}$ are also different. If both pairs, $A_{1}$ and $A_{2}$, $%
B_{1}$ and $B_{2}$ commuted, then the three probabilities would be the same,
as would the three final density \ operators - as can be seen using $%
\sum_{\gamma }\widehat{\Pi }_{\gamma }^{C}=1$ and $(\widehat{\Pi }_{\gamma
}^{C})^{2}=\widehat{\Pi }_{\gamma }^{C}$. In general however, the classical
probability $P(j,l|A_{1},B_{1})=\tsum%
\limits_{k,m}P(j,k,l,m|A_{1},A_{2},B_{1},B_{2})$ is linked to three quantum
probabilities for possible measurement processes that replicates \emph{one}
of the equivalent classical measurement processes on which $%
P(A_{1}=j,B_{1}=l)$ is based. \medskip \pagebreak

\end{document}